\documentclass[12pt]{article}
\pdfoutput=1 
\usepackage[utf8]{inputenc}
\usepackage{indentfirst}
\usepackage{amsmath}
\usepackage{graphicx}
\usepackage{float}
\usepackage{caption}
\usepackage{subcaption}
\usepackage{times}
\usepackage{authblk}
\usepackage[toc,page]{appendix}

\usepackage{amssymb}
\usepackage{amsbsy}
\usepackage{amsmath}
\usepackage{mathtools}
\usepackage[left, running, mathlines]{lineno}
 \usepackage{mathptmx}           
\usepackage{bm}   
\usepackage{framed}  
\numberwithin{equation}{section}

\newcommand{\bit}{\begin{itemize}}
\newcommand{\eit}{\end{itemize}}

\def\benu{\begin{enumerate}}
\def\eenu{\end{enumerate}}
\def\noi{\noindent}
\def\btab{\begin{tabbing}}
\def\etab{\end{tabbing}}

\def\bit{\begin{itemize}}
\def\eit{\end{itemize}}
\def\beq{\begin{equation}}
\def\eeq{\end{equation}}
\def\bec{\begin{center}}
\def\eec{\end{center}}
\def\btable{\begin{tabular}}
\def\etable{\end{tabular}}
\def\beqr{\begin{eqnarray}}
\def\eeqr{\end{eqnarray}}

\def\Rarw{\Rightarrow}

\def\eps{\epsilon}

\def\half{\frac{1}{2}}

\def\btab{\begin{tabbing}}
\def\etab{\end{tabbing}}
\def\beqrs{\begin{eqnarray*}}
\def\eeqrs{\end{eqnarray*}}
\def\noi{\noindent}

\def\bibi{\bibitem}
\def\bfig{\begin{figure}}
\def\efig{\end{figure}}
\def\fr{\frac}
\def\non{\nonumber}
\def\barr{\begin{array}}
\def\earr{\end{array}}

 \fontfamily{ptm}\selectfont

\title{\mbox{}\\   \mbox{} \\
Space Charge Driven Emittance Growth and the Effect of Octupoles in IOTA}

\author[2]{David Feigelson}
\author[1]{Tanaji Sen  \footnote{tsen@fnal.gov}}
\author[1]{Jean-Francois Ostiguy}
\author[3]{Runze Li}

\affil[1]{Fermi National Accelerator Laboratory, Batavia, IL 60510 }
\affil[2]{University of Chicago, Chicago, IL 60637 }
\affil[3]{ University of Wisconsin, Madison, WI 53706 }

 \date{}

\begin{document}

\maketitle

\begin{abstract}

The Integrable Optics Test Accelerator (IOTA) at Fermilab is 
a small machine dedicated to a broad frontier accelerator physics program. An important aspect of this program is to investigate the potential benefits of the resonance free 
tune spread achievable with integrable optics to store and accelerate high intensity proton beams for which space charge is significant. In this context, a good understanding of proton beam emittance growth and particle loss mechanisms is essential. 

Assuming nominal design parameters, simulations show that for a bunched beam, the bulk of emittance growth takes
place immediately following injection, typically within tens of turns. We attempt to account for this growth using a
simplified RMS mismatch theory; some of its limitations and possible improvements are briefly discussed. We then
compare  theoretical predictions to simulations performed using the PIC code pyORBIT. Further exploring ways to
mitigate emittance growth and reduce particle loss, we compare two beam matching strategies: (1) matching at the
injection point (2) matching at the center of the nonlinear (octupole) insertion region where $\beta_x = \beta_y$. To
observe how nonlinearity affects emittance growth and whether it dominates growth due to mismatch, we track two
different distributions. Finally, we explore the potential of using octupoles in a quasi-integrable configuration to
mitigate growth using a variety of initial distributions both at reduced and full intensities.
\end{abstract}
\pagebreak
\tableofcontents
\label{sec: emit_loss}  
\pagebreak
\section{Introduction}
Understanding emittance growth mechanisms in proton beams is of high importance for the IOTA program. In this report we investigate such mechanisms, focusing on RMS mismatch. First, we provide an overview of previous work as well as
a brief summary of known and relevant sources of growth.

Previous work has shown that at full intensity, over a period of 1000 turns, RMS emittance grows 10-fold and particle losses exceeds $1\%$ \cite{R1}. The bulk of the emittance increase occurs in the first few turns; this suggests that it cannot be due to betatron resonances. The time scale is consistent with the plasma oscillation period in IOTA, which is approximately 0.1 turn for a 3D Gaussian bunch \cite{R1}. 

Some attempts have been made at mitigating growth and losses. In particular, matching of the RMS beam parameters at the injection point did not succeed in limiting emittance growth but did have a significant impact on losses, reducing the latter to about $0.2\%$ \cite{R1}. Slow initialization without RMS matching was also shown to be effective, reducing emittance growth by roughly a factor of 2 and particle losses to about $0.01\%$ \cite{R1}.

The evolution of a general distribution $f(q_i,p_i)$ of states $(q_i,p_i)$ in a particle beam is governed by the Vlasov equation, which is itself a simplified form of the Liouville equation.
 In an accelerator where the position $s$ along the reference orbit is used as the independent dynamical variable, the distribution $f(q_i,p_i,s)$ is stationary if  ${\partial f}/{\partial s} = 0$.

An important observation is that a beam distribution $f(q_i,p_i,s)$ that can be expressed as a function of a dynamical invariant $W$ --- $f(W(q_i,p_i,s))$ --- is stationary. This is readily demonstrated as follows:  
By the definition of a dynamical invariant, $dW/ds = 0$.
Substituting into Liouville's equation $df/ds = 0$ for the distribution function  yields immediately
\begin{equation}
	0 = \frac{df}{ds} =  \frac{df}{dW}\fr{dW}{ds}  + \frac{\partial f}{\partial s}
\end{equation}
which implies
\begin{equation}
\frac{\partial f}{\partial s} = 0
\end{equation}
For an uncoupled accelerator, in the absence of space charge, a suitable invariant for generating a stationary distribution is the Courant-Snyder invariant 
\begin{equation}
W_x(x,x',s) = \gamma(s) x^2 + 2\alpha(s) xx' + \beta(s) x^2
\end{equation}
A Gaussian distribution of the form 
\begin{equation}
f(x,x',y,y',s) = A \exp \left\{ - \left[\frac{W_x}{2\epsilon_x} +\frac{W_y}{2\epsilon_y} \right] \right\}	
\end{equation}
would therefore be stationary in a linear lattice. In such a lattice, 
passage through an element transforms a quadratic form into another quadratic form. Furthermore, due to symplecticity, the determinant of the matrix of second moments i.e. the rms emittance is preserved so it would appear that rms emittance remains constant.
In practice, due to the unavoidable presence of residual non-linearity and chromatic effects, the phases of the betatron motion of individual particles advance at  sightly different rates. As a result, the particles eventually fill an elliptical phase region whose boundary is defined by the outermost particle in the initial phase space distribution. The phase space contours defined by the machine invariant are referred to as the matched contours.    

We conclude that a Gaussian distribution matched to the invariant contours is stationary in a linear lattice; however, an unmatched Gaussian will evolve into a broader distribution, leading to emittance  growth.

Recall that the net force acting on any particle is the sum of external forces provided by magnets and self forces due to particle interactions e.g. due to space-charge.
In the presence of space charge, there are very few examples of analytical self-consistent stationary distributions. A notable one is the so-called Kaptchinsky-Vladimirsky (K-V) distribution, which strictly speaking is defined in 4D phase space (i.e. it applies to an infinitely long uniform beam) as follows:
\begin{equation}
f(x.x',y.y') \propto \delta(1- \frac{W_x(x,x')}{\epsilon_{0x}} - \frac{W_y(y,y')}{\epsilon_{0y}}) 
\end{equation}
and $\epsilon_{0x}, \epsilon_{0y}$ are constants. It can be shown that the projection of such a distribution in physical $x-y$ space, i.e. the charge density $\rho(x,y)$, is uniform over an elliptical region. It follows that the corresponding space charge field is a linear function of the coordinates and therefore a properly matched K-V distribution in constant focusing channel is self-consistent and stationary. 

For general charge distributions of elliptical symmetry of the form $\rho = \rho(\frac{x^2}{a^2} + \frac{y^2}{b^2},s)$, an important result was  obtained by Sacharer \cite{Sacharer}: the linear part of the self-field depends mainly on the rms size of a distribution and only weakly on its exact form.
Moreover, the second moments obey the same envelope equation
as the local K-V distribution via the concept of rms equivalence.  

This suggests that a strategy to achieve stationarity in the presence of space charge is to match the rms beam moments to the lattice. Specifically, a beam is rms matched when its rms size remains constant as outlined in the following section. If two distributions have comparable rms sizes and experience similar growth, one may conclude that size mismatch is a more important factor than the details of the beam distribution in determining the emittance increase.
 
\section{RMS Mismatch}
\label{sec: mismatch}

The envelope equations obtained by Sacharer for a general elliptic beam are as follows:
\begin{eqnarray}
\tilde{x}'' + k^2_x(s)\tilde{x} - \frac{\epsilon_x^2}{\tilde{x}^3} - \frac{2 K_\text{sc}}{\tilde{x}+\tilde{y}} & = & 0 \\
\tilde{y}'' + k^2_y(s)\tilde{y} - \frac{\epsilon_y^2}{\tilde{y}^3} - \frac{2 K_\text{sc}}{\tilde{x}+\tilde{y}} & = & 0 
\end{eqnarray}
where $\tilde{x}$ and $\tilde{y}$ are the rms values of the distribution coordinates, 
\beq
K_{sc} = \frac{e\lambda_L}{2\pi\epsilon_0\beta\gamma^2pc}
\eeq
is the generalized dimensionless perveance, $\lambda_L$ is the charge line density and $k_x(s)$, $k_y(s)$ are respectively the horizontal and vertical lattice focusing strengths.  
For simplicity, we assume equal focusing in $x$ and $y$ and neglect the dependence of the focusing on $s$; this is known as the smooth focusing approximation and is generally a good approximation for FODO lattices with phase advance per cell less than $90^\circ$. The 2D envelope equations then reduce to the 1D form
\beq
a'' + k_0^2a - \frac{K_{sc}}{a} - \frac{\epsilon^2}{a^3} = 0
\label{envelope-equation}
\eeq
where the effective radius $a = \tilde{x} + \tilde{y} = 2 \tilde{x}$ and $k_0 = \nu_0/R$ is the 
uniform linear focusing strength, $\nu_0$ is the bare lattice tune while $R$ is the machine radius.  
A stationary beam in a linear focusing channel is characterized by a constant effective radius, so $a'' = 0$. This state corresponds to perfect balance between the external focusing force, the space charge force and the emittance divergence term. Then
\beq
k_0^2a - \frac{K_{sc}}{a} - \frac{\epsilon^2}{a^3} = 0
\label{equilibrium}
\eeq
Due to increased field energy, the emittance of a mismatched beam will evolve. If the space charge (second term in (\ref{equilibrium})) term dominates the emittance term (third term in \ref{equilibrium}), the beam transverse size can grow without bound. The transition occurs at 
\beq
K_{sc}a^2 = \epsilon^2
\eeq
If $K_{sc}a^2 > \epsilon^2$, growth is space charge dominated.
We have $\eps = k a^2 $ where $k$ is the net focusing strength after accounting for space charge.
\begin{equation}
 k^2 = k_0^2 - \frac{K_\text{sc}}{a^2}
\end{equation}
Defining $\nu$ as the tune in the presence of space charge, with $\nu< \nu_0$, 
the tune depression $\nu/\nu_0$ at the transition is found using
  \beqr
K_{sc} & = & k^2 a^2 \Rarw a^2(k_0^2 - k^2)   = k^2 a^2 \non \\
\fr{\nu}{\nu_0} =  \fr{k}{k_0} &  = &   \sqrt{\half} \approx 0.707
  \eeqr
\begin{figure}[H]
\centering
\includegraphics[scale = 0.6]{ 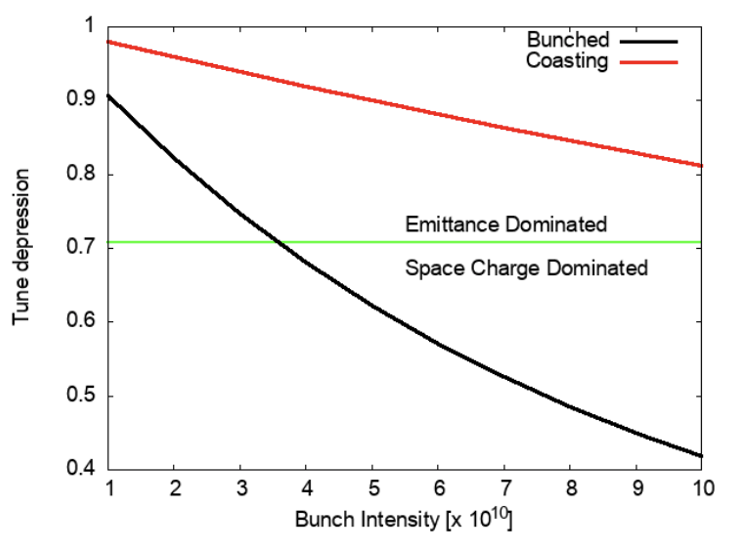}
\caption{Transition from emittance dominated to space charge dominated regimes for bunched and coasting beams in IOTA}
\label{fig: tune_depression}
\end{figure}
Figure \ref{fig: tune_depression} shows the tune depression as a function of bunch intensity for both bunched and coasting beams. Even at full design intensity ($10^{11}$), a coasting beam in IOTA is not space charge dominated \cite{R2}. This is reflected in previous work which showed substantially less emittance growth for a coasting beam than for a bunched beam \cite{R1}. It can also be seen that for a bunched beam, transition to the space charge dominated regime occurs at an intensity of $4 \times 10^{10}$. Beyond that threshold, the emittance is expected to grow substantially.
Using the simple picture of a uniform equivalent beam, Reiser \cite{Reiser} has shown how one can obtain a quantitative prediction of emittance growth arising from an imperfectly matched injected beam. We summarize the essential results.  

Let the total energy per particle in the mismatched beam be $E_n$ and the energy per particle in the equivalent matched beam be $E_i$.  The difference $\Delta E = E_n-E_i$ represents free energy that can be converted into thermal motion, i.e. emittance growth.
One can show that for a uniform stationary beam the total energy per particle, that is the sum of the kinetic energy, the potential energy due to external focusing, and the self field energy due to space charge 
\begin{eqnarray}
E & = & E_k + E_p + E_s\\
  &  =&   \frac{\gamma m v^2}{4} \left[ \left[k a^2 + k_0 a^2\right] + \frac{1}{2}[k_0-k^2] a^2 (1+ 4\ln \frac{b}{a}) \right]
\end{eqnarray}
where $a$ is the equivalent rms size, $b$ is the (conducting) beam pipe radius.
If we assume that the imperfectly matched beam relaxes into a matched state, the energy in the final stationary state must be equal to the energy in the initial state. With $a_i$, $a_f$ the initial and final rms equivalent sizes of the beam and $k_i$,$k_f$ the initial and final net focusing constants
one has 
\begin{multline}
\frac{\gamma m v^2}{4} \left[ \left[k_f a_f^2 + k_0^2 a_f ^2\right] + \frac{1}{2}[k_0-k_f^2] a_f^2 (1+ 4\ln \frac{b}{a_f}) \right] = \\ 	
\frac{\gamma m v^2}{4} \left[ \left[k_f a_f^2 + k_0^2 a_f ^2\right] + \frac{1}{2}[k_0-k_f^2] a_f^2 (1+ 4\ln \frac{b}{a_f}) \right] + \Delta E
\label{energybalance}
\end{multline}
The free energy may be expressed in the form
\begin{equation} 
\Delta E = \frac{1}{2} \gamma mv^2 k_0^2 a_i^2 h
\label{freeenergy}
\end{equation}
where $h$, the ratio between the actual and the equivalent uniform distribution energy per particle, is a
dimensionless factor or order unity (referred to as the free energy parameter) that can  be calculated. 
Using (\ref{energybalance}) and (\ref{freeenergy}) we obtain a relation between $a_f$ and $a_i$ 
\begin{equation}
\left(\frac{a_f}{a_i}\right)^2 - (1-\frac{k^2_i}{k^2_0})\ln \frac{a_i}{a_f} = h  
\end{equation}	   
which for $a_f-a_i << a_i$ (usually true with moderate space charge effects) may be simplified to
\begin{equation}
	\frac{a_f}{a_i} \simeq 1 + \frac{h}{1+(k_i/k_0)^2}
\label{sizeratio}
\end{equation}	
At high intensities, as is the case in IOTA at full intensity, there is large emittance and beam size growth for which
the above approximation is not valid. 
Finally, from the equilibrium condition (\ref{equilibrium}) we get the difference in emittance between the final and initial stationary beams
\begin{equation}
\Delta \epsilon^2 = \epsilon_f^2-\epsilon_i^2 = k_f^2 a^4_f -k_i^2 a_i^4	
\end{equation}
The complete set of  equations for calculating emittance growth without further approximations  is \cite{R2},
\beqr
k_i^2 a_i^4 &  = &  \eps^2, \;\;\;    k_i^2 \equiv  k_0^2 - \fr{K_{sc}}{a_i^2}  \label{eq: eqs_set_1}  \\
h & =  & \half \fr{k_i^2}{k_0^2} [ \fr{a_i^2}{a_0^2} - 1 +
  \fr{k_0^2}{k_i^2}( \fr{a_0^2}{a_i^2} - 1) +  2 \fr{K_{sc}}{k_i^2 a_i^2}\ln(\fr{a_i}{a_0}) ]
\label{eq: eqs_set_3} \\
h & = &   \fr{a_f^2}{a_i^2} - 1 - [1 - \fr{k_i^2}{k_0^2} ] \ln (\fr{a_f}{a_i})  \label{eq: eqs_set_4} \\
\fr{\eps_f}{\eps_i} & = &  \fr{a_f}{a_i}\left[ 1 + \fr{k_0^2}{k_i^2}\{(\fr{a_f}{a_i})^2 - 1 \} \right]^{1/2}  \label{eq: emit_f_i} \label{eq: eqs_set_5}
\eeqr
The procedure to use them is as follows:
\benu
\item Given the tune, initial emittance and bunch intensity, solve the envelope equation (\ref{eq: eqs_set_1})
  to find the initial matched beam size  $a_i$.
\item  Use Eq.(\ref{eq: eqs_set_3}) to find $h$.
\item Use Eq.(\ref{eq: eqs_set_4}) to find the ratio of final to initial matched beam sizes $a_f/a_i$
\item Use Eq.(\ref{eq: eqs_set_5}) to find the ratio of final to initial emittances $\eps_f/\eps_i$
  \eenu

\subsection{Theory vs Simulations}
\label{sec: theory_sim}
\begin{figure}[H]
\centering
\includegraphics[scale = 0.39]{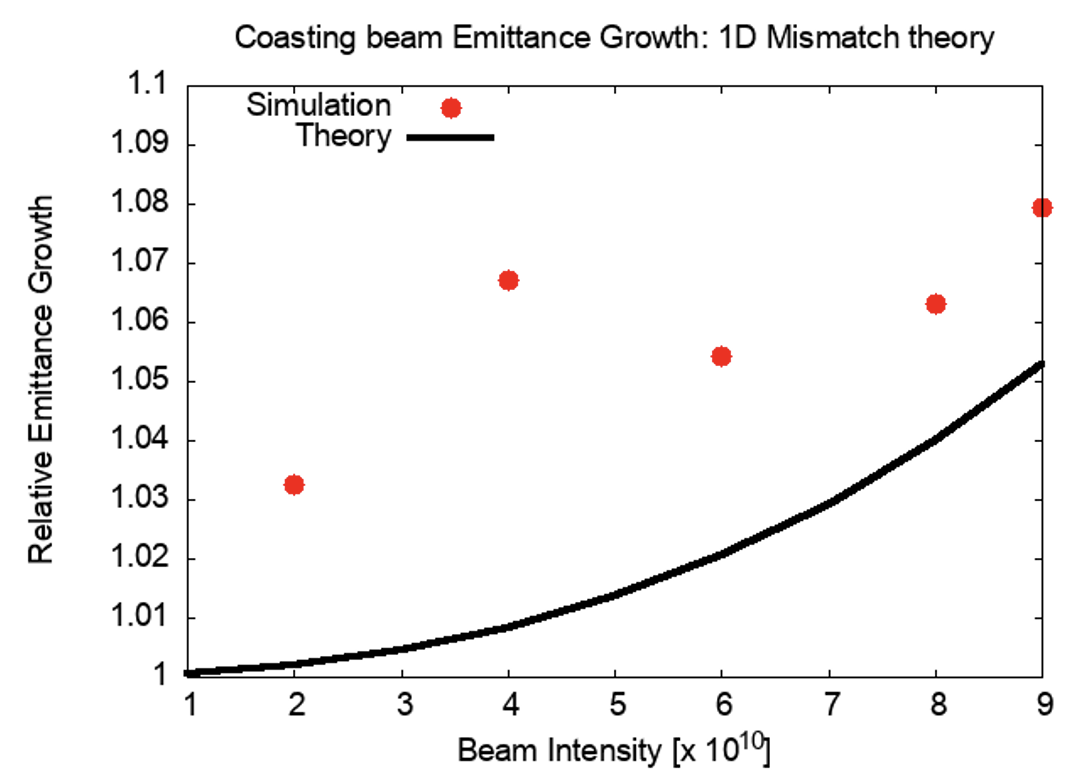}
\includegraphics[scale = 0.35]{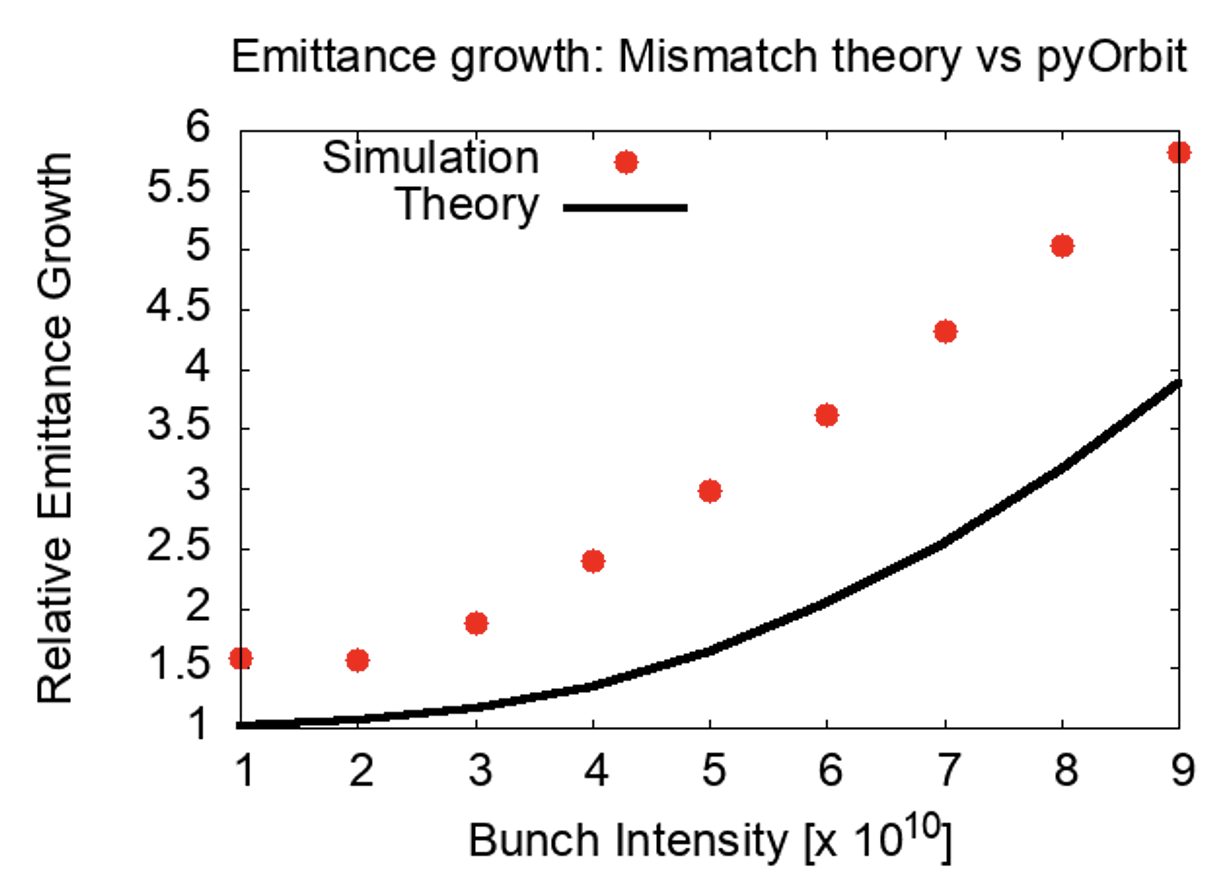}
\caption{Growth predicted by 1D mismatch theory and pyORBIT for coasting (left) and bunched beams (right)}
\label{fig: theory_vs_simulations}
\end{figure}

Though the search for better theoretical models is ongoing, the simple 1D mismatch  theory can already account for more than half of emittance growth observed in pyORBIT simulations. Figure \ref{fig: theory_vs_simulations} compares the theory to pyORBIT simulations for both bunched and coasting beams.

For a coasting beam, the disparity between theory and simulation is about $2\%$ at full intensity. For a bunched beam at full intensity, the theory can still account for about 2/3 of a 6-fold increase.

It is important to reiterate that the simplified theory assumes the smooth focusing approximation holds and does not take dispersion or transverse coupling into consideration. That said, it is encouraging that despite the relative crudeness of the theory, predictions are generally compatible with simulations results. 

\subsection{RMS Matching at a high symmetry location (\boldsymbol{$\beta_{x} = \beta_{y}$})}

\begin{figure}[H]
\centering
\includegraphics[scale = 0.3]{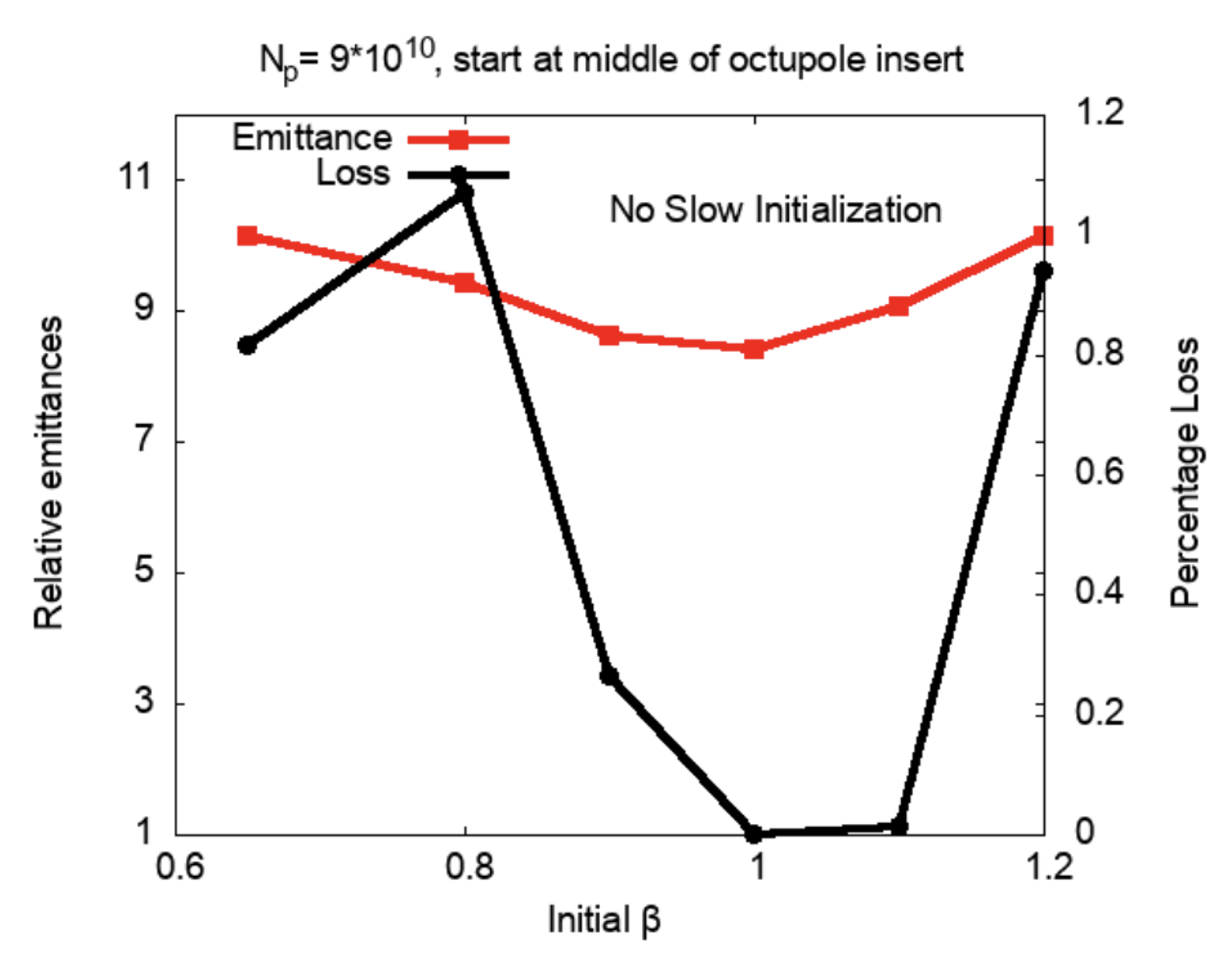}
\caption{Emittance growth and particle loss as a function of the initial matched $\beta$.}
\label{fig: rms_new}
\end{figure}

In the presence of space-charge, RMS matching is expected to be an effective strategy to achieve reasonably stationary conditions and therefore control emittance growth and particle loss. In principle, matching can be performed at any location in the lattice; however, in practice it is simplest to perform matching at a location of high symmetry. We choose the center of the octupole insert, where (in the absence of space charge) $\beta_{x} = \beta_{y}$, $\alpha_{x} = \alpha_{y} = 0$, and $D_x = D = 0$. The impact of this matching on emittance growth and particle loss at full intensity is shown in Figure \ref{fig: rms_new}; no slow initialization was performed. At an optimal value of $\beta$, loss is reduced to $0\%$, even lower than the $0.2\%$ achieved with the previous match at the normal injection point. That said, emittance growth remains significant, increasing by a factor of $8$ to $10$ from its initial value. These results reaffirm the observation that while RMS matching is helpful in reducing loss, it has a much lesser impact on emittance growth. An explanation for this state of affairs is a topic for future work.

\section{Non-Equilibrium Distributions }

While mismatch is a significant source of emittance growth, non-equilibrium distributions also have an impact. To better understand this source of growth, we compared the  Gaussian and a flat-topped  distribution in
bunched beams.  The latter distribution  has a uniform charge density, but over a region delimited by a rectangular
boundary rather than an elliptic one (for a KV distribution). While the space charge field for such a distribution is linear
in the vicinity of its center, significant deviations from linearity may be expected near the boundary, in the corner
regions. Comparisons with a KV distribution will be discussed in a future report. 

\begin{figure}[H]
\centering
\includegraphics[scale = 0.6]{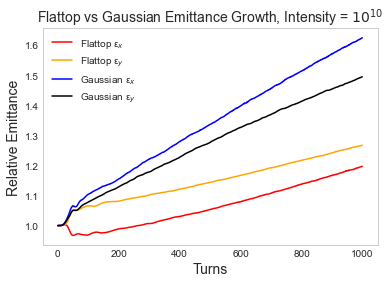}
\caption{Relative emittance growth for flat-topped and Gaussian distributions  at a bunch intensity of
  1$\times 10^{10}$.}
\label{fig: kv_gs_1e10}
\end{figure}

Figure \ref{fig: kv_gs_1e10} shows the discrepancy in growth between the two distributions at an intensity of
$10^{10}$ with 40 turns of slow initialization and matched initial emittance for a Gaussian and a 
flat-topped distribution. There is no particle loss over 1000 turns in either case.
Emittance growth for the Gaussian distribution is greater than for the flat-topped, increasing by a factor 1.6 in $x$
and 1.5 in $y$ vs 1.2 in $x$ and 1.3 in $y$ respectively. This confirms that the specifics of a non-equilibrium
distribution have an  impact on emittance growth at low intensity.

\begin{figure}[H]
\centering
\includegraphics[scale = 0.46]{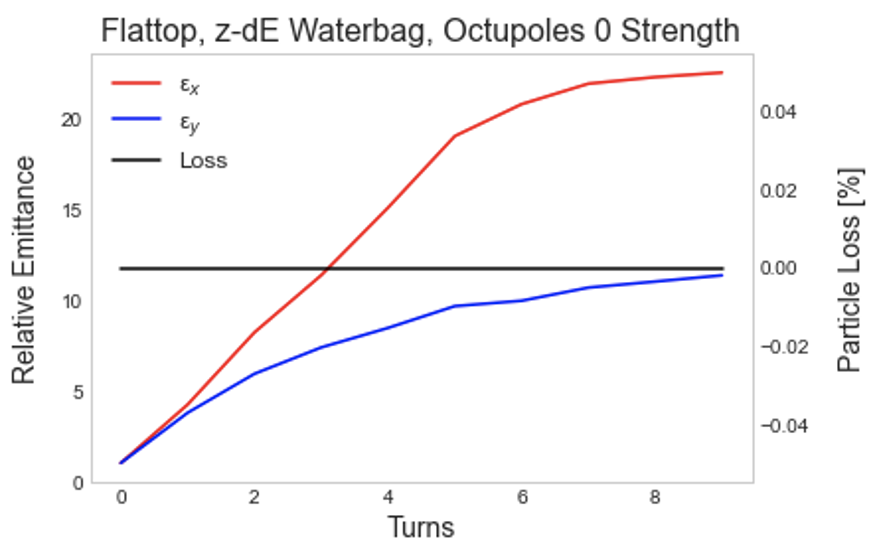}
\includegraphics[scale = 0.46]{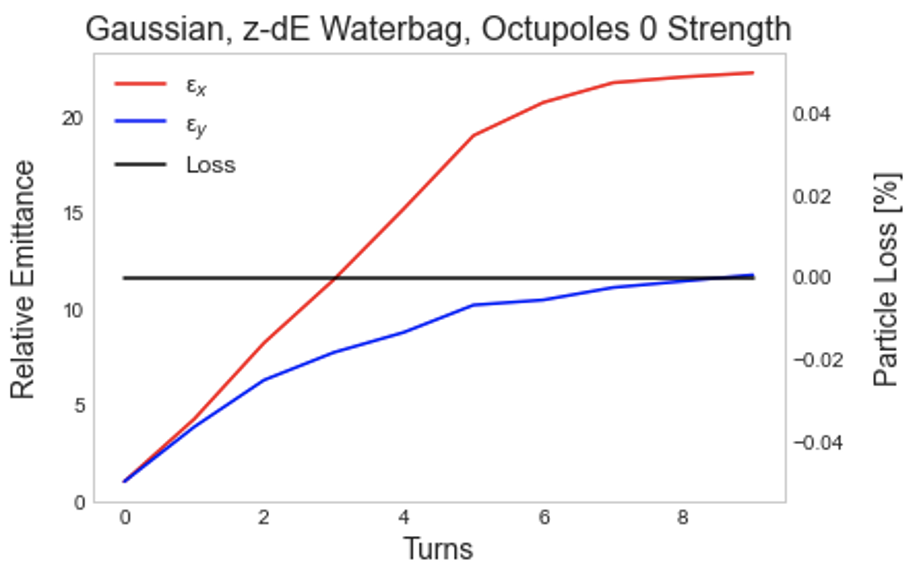}
\caption{Emittance growth and losses over the first 10 turns for flat-topped (left)  and Gaussian  distributions
  (right) at a bunch intensity of 9$\times 10^{10}$.}
\label{fig: 9e10_init_kvgs}
\end{figure}

At full intensity, growth occurs rapidly, on a much shorter timescale, possibly due to a different dominant source of
growth. With no slow initialization, and truncated initial distributions with matched initial emittance, emittance growth
is very similar between the distributions over the first ten turns, as shown in Figure \ref{fig: 9e10_init_kvgs}. In both cases, emittance grows by a factor 20 horizontally and 10 vertically.

\begin{figure}[H]
\centering
\includegraphics[scale = 0.55]{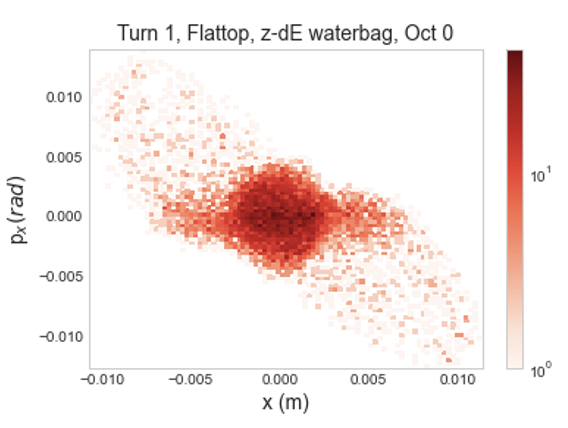}
\includegraphics[scale = 0.55]{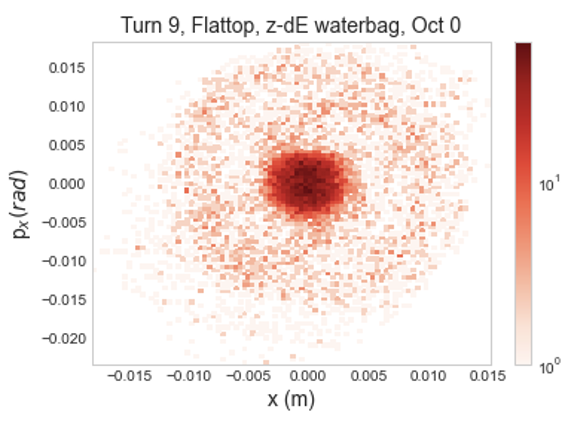}
\includegraphics[scale = 0.55]{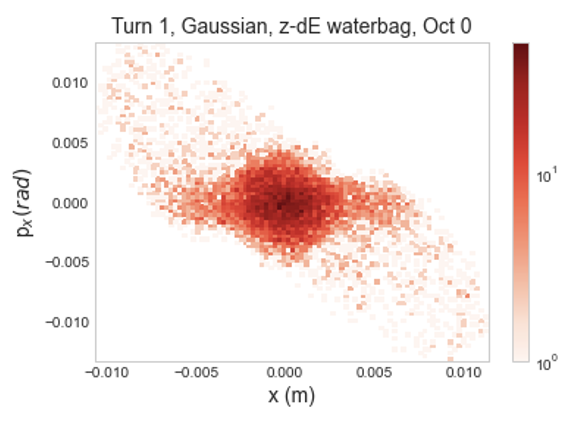}
\includegraphics[scale = 0.55]{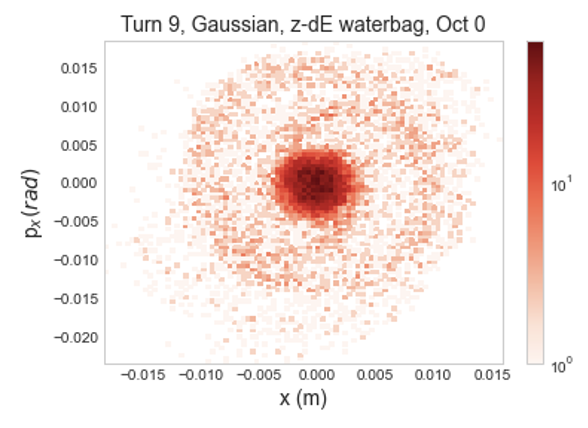}
\caption{Phase space growth in x after the first 10 turns for the flat-topped  distribution (top row) and for the
  Gaussian distribution (bottom row). }
\label{fig: 2Dhist_x}
%
\centering
\includegraphics[scale = 0.55]{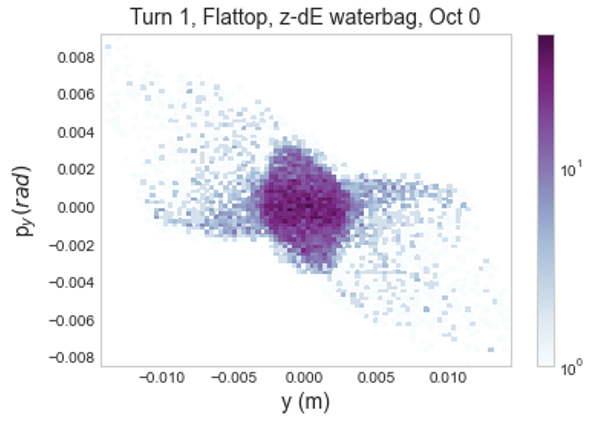}
\includegraphics[scale = 0.55]{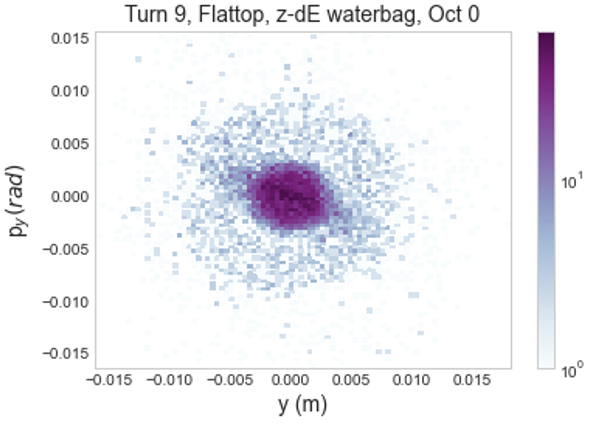}
\includegraphics[scale = 0.55]{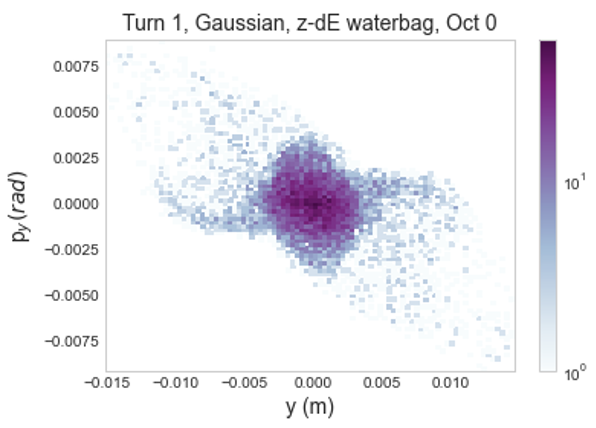}
\includegraphics[scale = 0.55]{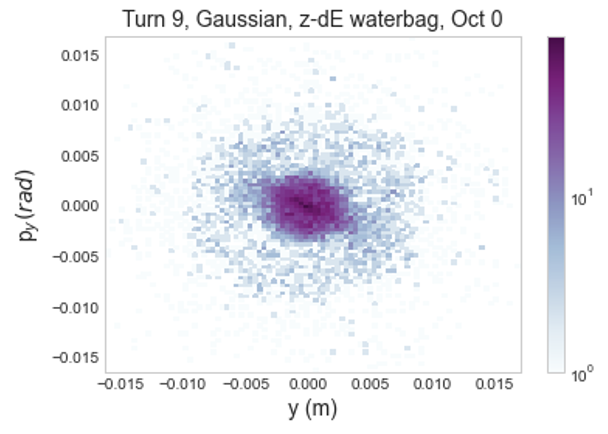}
\caption{Phase space growth in y after the first 10 turns for the flat-topped  distribution (top row) and for the
  Gaussian distribution (bottom row).}
\label{fig: 2Dhist_y}
\end{figure}

\begin{figure}[H]
\centering
\includegraphics[scale = 0.5]{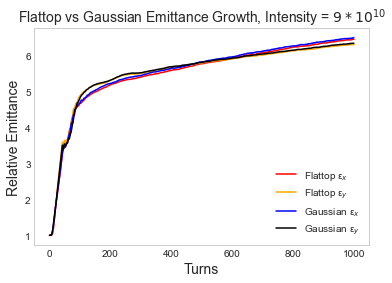}
\includegraphics[scale = 0.5]{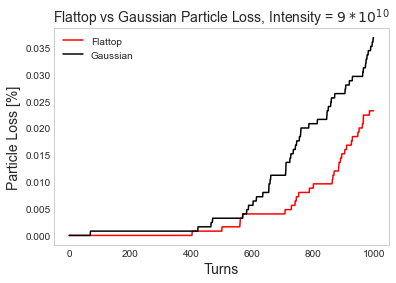}   
\caption{Emittance growth and losses over 1000 turns for flat-topped (left) and Gaussian (right) distributions
  at a bunch intensity of 9$\times 10^{10}$.}
\label{fig: kv_gs_9e10}
\end{figure}

Examining this growth in 2D histograms, shown in Figures \ref{fig: 2Dhist_x} and \ref{fig: 2Dhist_y}, there is very
similar initial behavior between the two distributions. Horizontal phase space rapidly develops a populated halo,
extending to 0.015 m in $x$ and 0.015 rad in $x'$. Halo growth in the vertical plane is substantially more constrained, extending to roughly 0.01 m in $y$ and 0.01 rad in $y'$.
Looking at longer term growth i.e. over 1000 turns instead of the initial 10, the two distributions still behave very similarly. Figure \ref{fig: kv_gs_9e10} shows this for 40 turns of slow initialization and a full Gaussian distribution and a flat-topped distribution with matched initial emittance. This suggests that mismatch, rather than the specifics of the distribution, may be a stronger source of growth at full intensity. This is interesting as it hints at a change in the dominant source of growth as intensity increases. There is also notably more loss with the Gaussian distribution, at about $0.035\%$ as compared to about $0.025\%$. This suggests greater halo growth over 1000 turns.

\section{Space Charge and Octupoles}

An important aspect of research at IOTA is the potential for either quasi or fully integrable optics to mitigate space charge resonances. We begin this investigation by looking at the impact of an octupole string insert (quasi-integrable optics) on beam dynamics. Previous work found that the dynamic aperture in the presence of this type of insert is reduced to approximately $3\sigma$ \cite{R1}. The impact of such a reduction is evident in many of the simulation results.

In all simulations, particles are tracked for 1000 turns. To allow the beam to relax into any physically accessible steady
state, the physical aperture radius is set to an arbitrarily large value (10 cm). We examine a bunched beam with reduced and full beam intensities, octupoles at various strengths, and selected initial particle distributions.

\subsection{Low Intensity Results}
\label{low_intensity}

\begin{figure}[H]
\centering
\includegraphics[scale = 0.3]{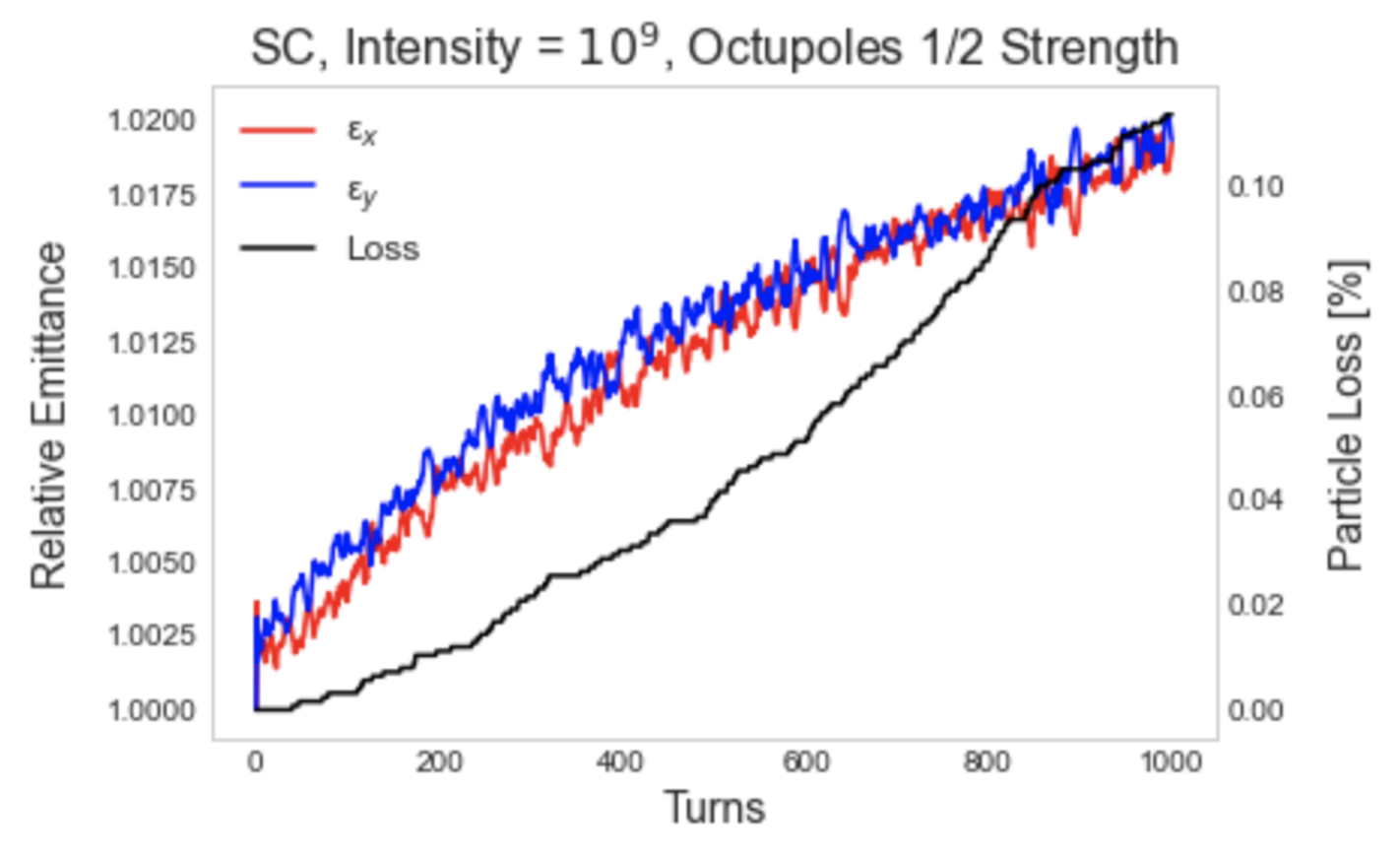}
\includegraphics[scale = 0.3]{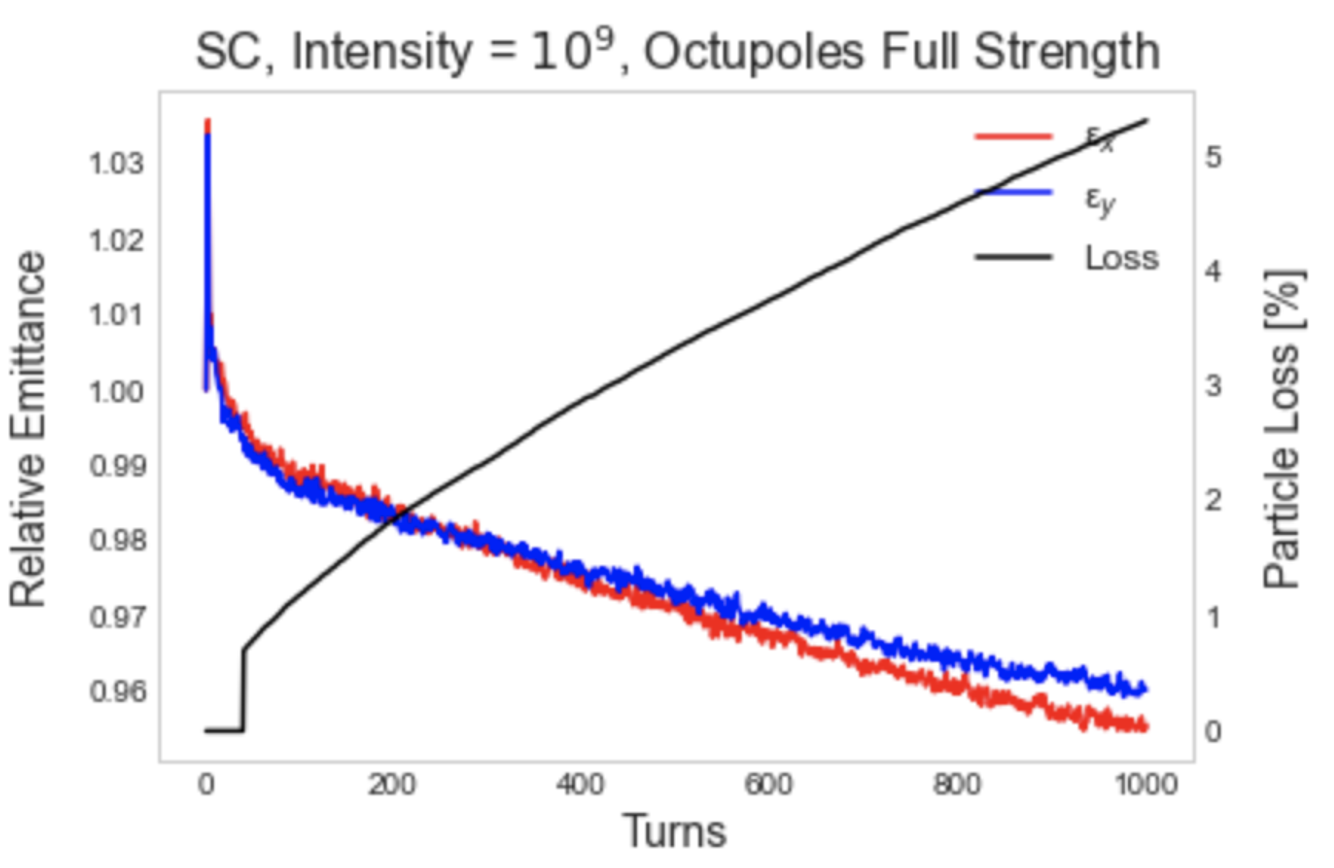}
\caption{Emittance growth and loss over 1000 turns with a full Gaussian distribution at intensity $10^9$.}
\label{fig: 1e9_growth}
\end{figure}

Even at a very low intensity of $10^9$, the impact of the decreased dynamic aperture is evident. With a full Gaussian distribution and full strength octupoles, particles almost immediately reach the dynamic aperture boundary. This is shown in Figure \ref{fig: 1e9_growth}, comparing a bunched beam with octupoles at $1/2$ strength and full strength.

At $1/2$ strength, particles remain within the dynamic aperture and emittance grows about $2\%$ with loss slightly
above $0.10\%$ after 1000 turns. However, with octupoles at full strength, particles immediately move outside of the dynamic aperture, as evidenced by the emittance spike right after turn 0. This induces substantial particle loss --- more than $5\%$ after 1000 turns.

\begin{figure}[H]
\centering
\includegraphics[scale = 0.39]{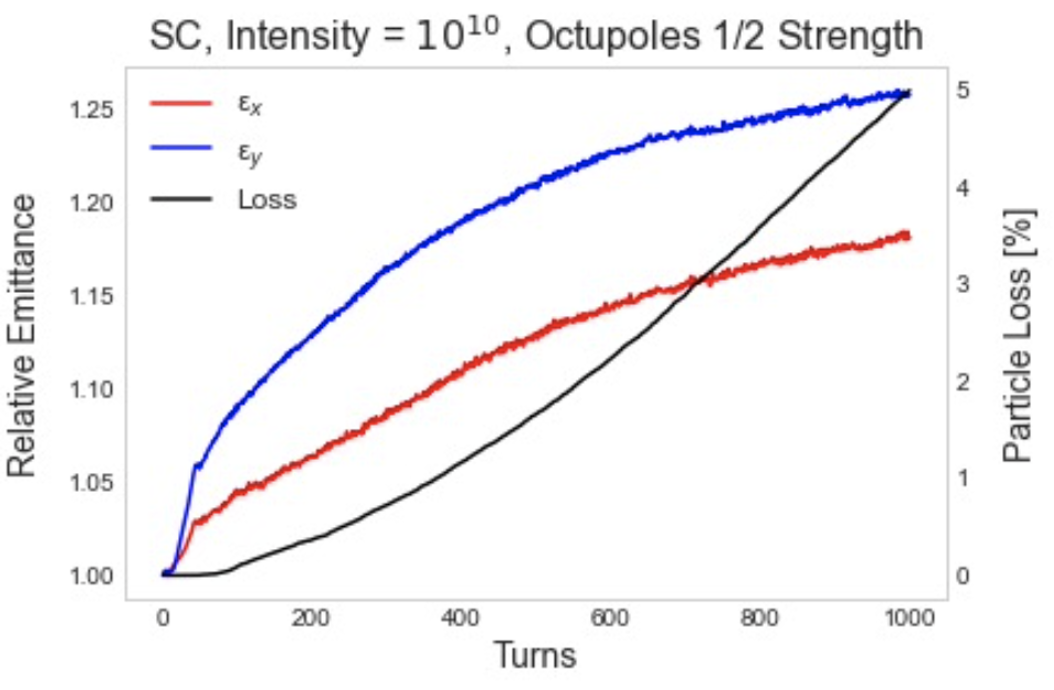}
\includegraphics[scale = 0.39]{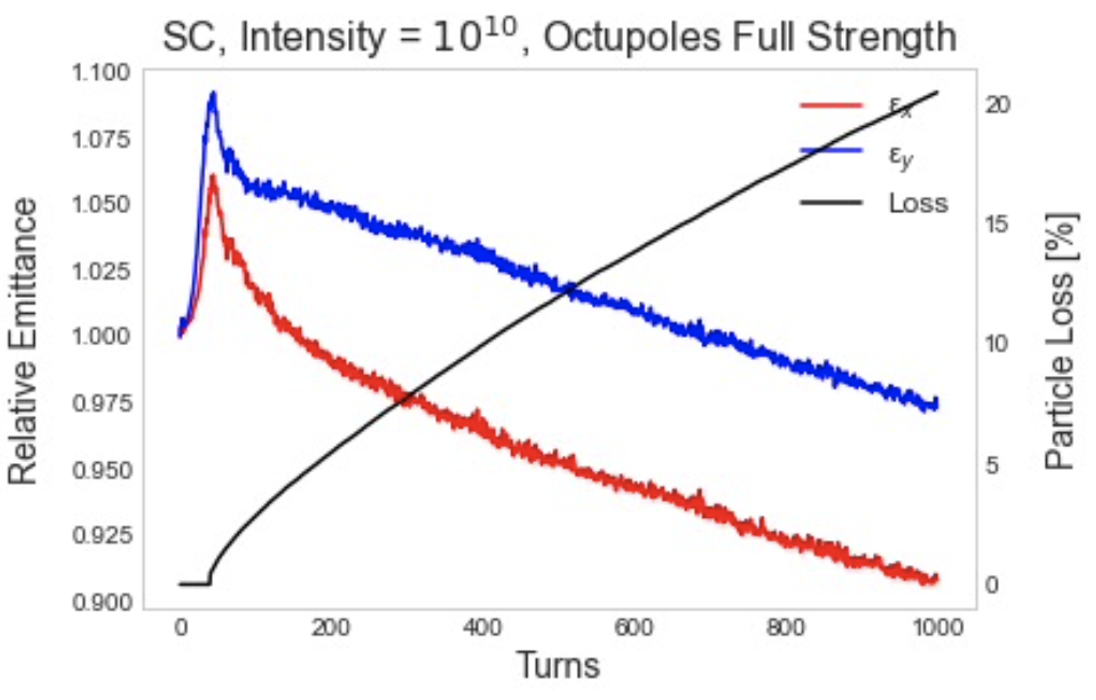}
\caption{Emittance growth and loss over 1000 turns with a full Gaussian distribution at intensity $10^{10}$.}
\label{fig: 1e10_growth}
\end{figure}

At a  higher intensity of $10^{10}$, the effect becomes more pronounced. Figure \ref{fig: 1e10_growth} shows the same comparison as before but at the increased intensity.  At half strength octupoles, there is still no immediate loss due to
the dynamic aperture but instead grows slowly to $\tilde 5\%$ over the 1000 turns. At full strength octupoles, there is
again immediate loss which grows more rapidly and exceeds $20\%$ over the same time. 

There is also an evident dispersion effect. At both octupole strengths, emittance growth is significantly larger in the
vertical plane. For example, for the $1/2$ strength case, $\epsilon_y$ grows $25\%$ while $\epsilon_x$ grows
$15\%$. The likely explanation is that dispersion increases the horizontal beam size, which causes a reduction of
horizontal space charge forces and therefore less emittance growth. The effect of dispersion is also observed in loss
distributions as a function of the transverse distance; at the same distances along $x$ and $y$, the loss in the $x$ plane
is greater. 

To prevent immediate loss due to the dynamic aperture, a truncated Gaussian distribution is effective. In the following simulations, the distribution was truncated to $1.5\sigma$. Without octupoles, such a distribution produces no loss over
1000 turns.

\begin{figure}[H]
\centering
\includegraphics[scale = 0.4]{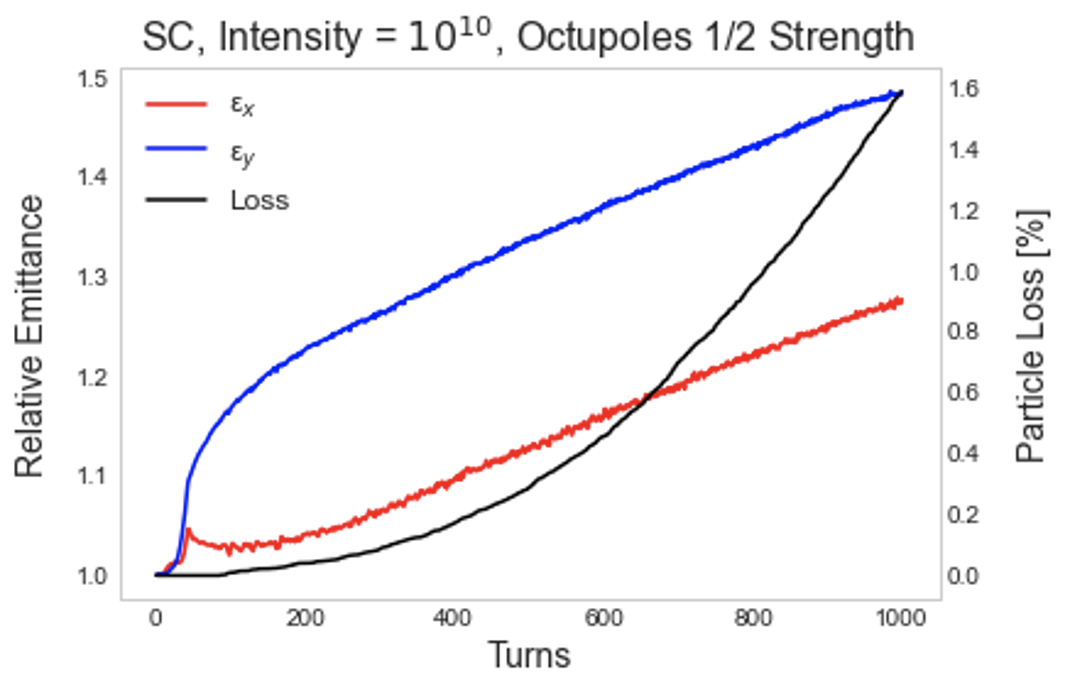}
\includegraphics[scale = 0.4]{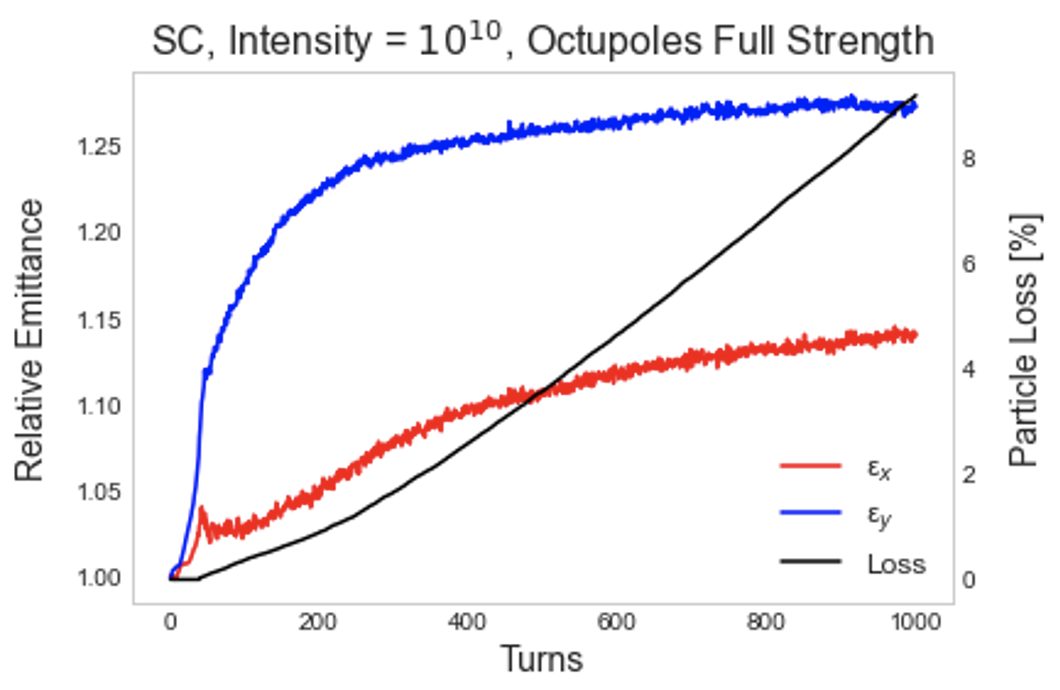}
\caption{Emittance growth and loss over 1000 turns with a truncated Gaussian distribution at intensity $10^{10}$.}
\label{fig: 1e10_trunc_growth}
\end{figure}

Figure \ref{fig: 1e10_trunc_growth} shows results from the truncated distribution. Notably, there is no immediate loss due to the dynamic aperture even with octupoles at full strength. Comparing these results with the full Gaussian distribution, we find that loss decreases from $5\%$ to $1.6\%$ with octupoles at half strength and from $20\%$ to $10\%$ with octupoles at full strength. $\epsilon_y$ still sees a greater increase than $\epsilon_x$, showcasing the same dispersion effect described above. There is a small peak early in $\epsilon_x$ which likely corresponds to halo loss. Again, the likely explanation is that the halo particle amplitudes are greater in the horizontal plane due to dispersion.

\subsection{Full Intensity Results}
\label{full_intensity}

\begin{figure}[H]
\centering
\includegraphics[scale = 0.4]{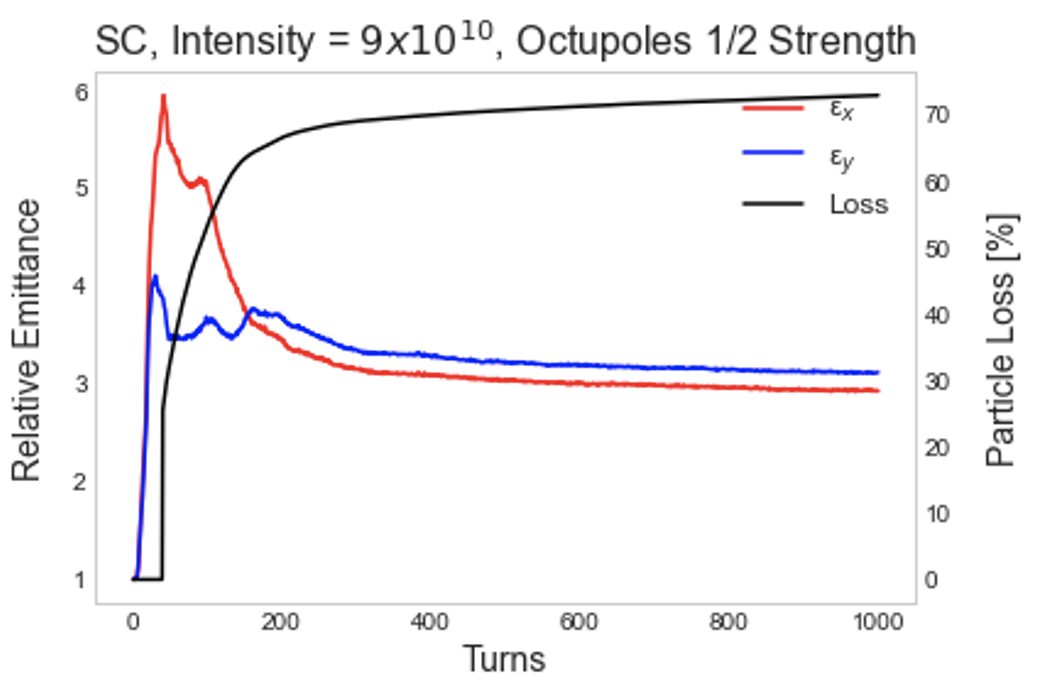}
\includegraphics[scale = 0.4]{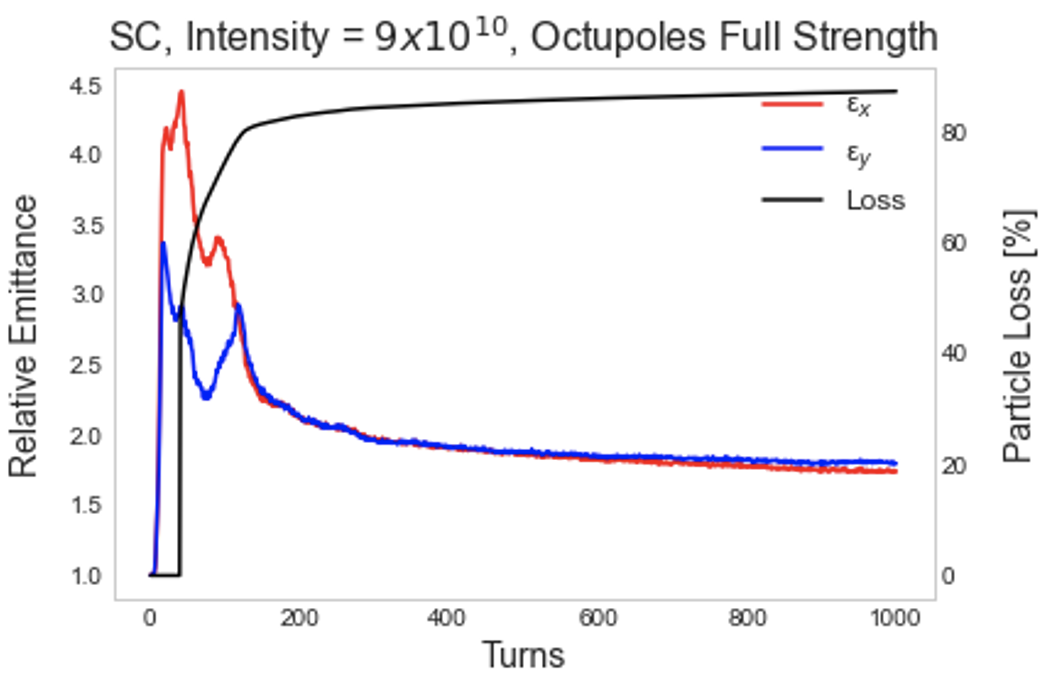}
\caption{Emittance growth and loss with a truncated Gaussian distribution at full intensity.}
\label{fig: 9e10_trunc_growth}
\end{figure}

At full intensity, with octupoles at both half and full strengths, even with a truncated Gaussian distribution, particles hit the dynamic aperture. Shown in Figure \ref{fig: 9e10_trunc_growth}, the initial behavior is chaotic and results in losses upwards of $70\%$ and $80\%$ respectively. While these results do not appear promising for operation with an octupole insert, further investigations are necessary before one can reach definitive conclusions.

\section{Conclusions}

\bit

\item 
  Initial mismatch theory suggests that bunched beams in IOTA are space charge dominated at intensities greater than $4 \times 10^{10}$. In contrast, the same theory suggests that coasting beams in IOTA remain emittance dominated through the maximum beam intensity of $9 \times 10^{10}$.

\item 
  Simulations show that RMS matching is effective at reducing beam loss. At full intensity of $9 \times 10^{10}$, loss over 1000 turns decreased from greater than $1\%$ in the linear lattice not RMS matched to space charge lattice functions to approximately $0.2\%$ in a lattice RMS matched at the injection point. RMS matching in the center of the octupole insert where $\beta_{x} = \beta_{y}$ and initial dispersion is $0$ showed even greater improvement, reducing loss to $0\%$. In both cases, emittance growth was not meaningfully reduced, changing from a roughly 9-fold increase to an 8-fold increase.

\item 
  At a reduced intensity of $10^{10}$, using a lattice RMS matched at the center of the octupole insert, and using 40 turns of slow initialization, emittance growth over 1000 turns for a Gaussian distribution is greater than that of a flat-topped distribution, increasing by approximately 1.6 compared to 1.2 horizontally and approximately 1.5 compared to 1.3 vertically. This shows that a non-equilibrium distribution has a significant impact on emittance growth at low intensity. On the other hand, with the same conditions at full intensity, emittance growth for the two distributions is very similar, with both increasing more than 6-fold in both x and y. This suggests that at full intensity, RMS mismatch is a stronger source of emittance growth than a non-equilibrium distribution.

\item 
  The presence of an octupole insert greatly reduces the dynamic aperture to about $3\sigma$, necessitating a truncated initial distribution to prevent a high level of immediate loss. With full strength octupoles, at a lower intensity of $10^{10}$, a full Gaussian distribution leads to  particle loss upwards of $20\%$ after 1000 turns while a distribution truncated to $1.5\sigma$ results in less than half of this loss.

\item
   The impact of dispersion is evident in the differences in emittance growth and loss between the horizontal and vertical planes. At an intensity of $10^{10}$, using a lattice RMS matched to the injection point, and octupoles at half strength, a truncated Gaussian distribution sees vertical emittance growth of $50\%$ and horizontal growth of $25\%$ after 1000 turns. With the same conditions and octupoles at full strength, emittance grows $25\%$ in y and $15\%$ in x. Growth is greater in the vertical plane because  dispersion increases horizontal beam size, reducing horizontal space charge forces. Dispersion also increases horizontal amplitudes, increasing horizontal loss relative to vertical loss.
  \eit

  \noi   {\large \bf Acknowledgments} \\
  We thank the SIST internship program at Fermilab which enabled the participation of David Feigelson in this study.
Fermilab is operated by Fermi Research Alliance LL under DOE contract No. DE-AC02CH11359.

\end{document}